\documentclass[floatfix,showkeys ]{revtex4}
\usepackage{graphicx}
\usepackage{dcolumn}
\usepackage{bm}
\usepackage{enumerate}
\usepackage{tipa}

\usepackage{amsmath}
\usepackage{slashed}
\usepackage{graphicx}
\usepackage{subfigure}
\usepackage{float}
\usepackage{epstopdf}
\usepackage{color}
 \usepackage{feyn}
\definecolor{darkgreen}{rgb}{0,0.5,0}
\definecolor{purple}{rgb}{0.5,0,0.5}
\definecolor{nblue}{rgb}{0.0,0.0,0.50}
\definecolor{scarlet}{rgb}{1.0,0.2,0}
\usepackage[colorlinks=true,linkcolor=purple,citecolor=purple,urlcolor=blue]{hyperref}
\begin{document}

\def\no{\noindent}
\def\non{\nonumber\\}
\def\epsp#1#2{\varepsilon_{#1}\cdot p_{#2}}
\def\be{\begin{equation}}
\def\ee{\end{equation}}
\def\bea{\begin{eqnarray}}
\def\eea{\end{eqnarray}}
\def\bear{\begin{eqnarray}}
\def\ear{\end{eqnarray}}
\def\bei{\begin{itemize}}
\def\eei{\end{itemize}}
\def\bee{\begin{enumerate}}
\def\eee{\end{enumerate}}
\def\noN{\nonumber}
\def\ccr{\nonumber\\}
\def\eqa{\!&=&\!}
\def\Eins{{\mathchoice {\rm 1\mskip-4mu l} {\rm 1\mskip-4mu l}
{\rm 1\mskip-4.5mu l} {\rm 1\mskip-5mu l}}}
\def\Z{{\mathchoice {\hbox{$\sf\textstyle Z\kern-0.4em Z$}}
{\hbox{$\sf\textstyle Z\kern-0.4em Z$}}
{\hbox{$\sf\scriptstyle Z\kern-0.3em Z$}}
{\hbox{$\sf\scriptscriptstyle Z\kern-0.2em Z$}}}}
\def\e{\,{\rm e}}
\def\veps#1{\varepsilon_{#1}}
\def\eps#1{\epsilon_{#1}}
\def\kk#1#2{k_{#1}\cdot k_{#2}}
\def\pp#1#2{p_{#1}\cdot p_{#2}}
\def\intT{\int_0^\infty dT\, {\rm e}^{-m^2T}}
\def\matD{\mathcal{D}}
\def\t #1{\tau_{#1}}
\def\eTx{{\rm e}^{-\frac{1}{4}\int_0^Td\tau\dot{x}^2}}
\def\eTq{{\rm e}^{-\frac{1}{4}\int_0^Td\tau\dot{q}^2}}
\def\vaeps{\varepsilon}
\def\ddel{{}^\bullet\! \Delta}
\def\deld{\Delta^{\hskip -.5mm \bullet}}
\def\dddel{{}^{\bullet \bullet} \! \Delta}
\def\ddeld{{}^{\bullet}\! \Delta^{\hskip -.5mm \bullet}}
\def\deldd{\Delta^{\hskip -.5mm \bullet \bullet}}
\def\epsk#1#2{\varepsilon_{#1}\cdot k_{#2}}
\def\epseps#1#2{\varepsilon_{#1}\cdot\varepsilon_{#2}}
\def\Gd{\dot{G}}
\def\Gdd{\ddot{G}}

\title{Multiphoton amplitudes and generalized LKF transformation in scalar QED using the worldline formalism}
\author{\underline{Naser Ahmadiniaz}$^{a,}$\footnote{ahmadiniaz@ibs.re.kr (Talk given at the International Workshop on Strong Field Problems in Quantum Theory, June 6-11, 2016, Tomsk, Russia)}, Adnan Bashir$^b,$\footnote{adnan@ifm.umich.mx} and Christian Schubert$^{b},$\footnote{schubert@ifm.umich.mx}}
\affiliation{$a$ Center for Relativistic Laser Science,\\
 Institute for Basic Science,\\
 Gwangju 61005, Korea\\
$^b$ Instituto de F{{\'\i}}sica y Matem\'aticas, Universidad Michoacana de San Nicol\'as de Hidalgo\\
Apdo. Postal 2-82, C.P. 58040, Morelia, Michoacan, Mexico
}

\date{\today}

\begin{abstract}

We apply the worldline formalism to scalar quantum electrodynamics (QED) to find a Bern-Kosower type master formula for generalized Compton scattering,
on-shell and off-shell. Moreover, we use it to study the non-perturbative gauge parameter dependence of amplitudes in scalar QED and, as our main result, 
find a simple non-perturbative transformation rule under changes of this parameter in $x$-space in terms of conformal cross ratios. 
This generalizes the well-known Landau-Khalatnikov-Fradkin transformation (LKFT). We also exemplify how the LKFT works in perturbation theory.

\end{abstract}

\maketitle

\section{Introduction}

Although the one-loop correction to the matter - gauge boson vertex in gauge theory has been studied long ago, the full non-perturbative structure of this vertex has remained a challenge for decades, not only in quantum chromodynamics (QCD) but also for simpler cases as quantum electrodynamics (QED). A systematic study for spinor QED was initiated more than three decades ago by Ball and Chiu \cite{bach-80} and led to their well-known decomposition of the vertex into transversal and longitudinal parts. 
They also calculated the vertex at one-loop in Feynman gauge. Later, the same vertex was calculated in
Yennie-Fried gauge \cite{adkins-94}. The extension of the Ball-Chiu results to an arbitrary covariant gauge was carried out in \cite{Pennington-95}. 
In three dimensional spinor QED, the massive and massless vertex in an arbitrary  covariant gauge was obtained in \cite{adnan1,adnan2,adnan3}. 

Landau and Khalatnikov and independently Fradkin have derived a series of transformations (LKFT) for QED that transform the Green functions in a specific manner under a variation of the gauge \cite{LK,fradkin}.  Later these transformations were rederived by Johnson and Zumino by means of functional methods \cite{john-zum}.  
These transformations are non-perturbative and written in coordinate space. They can not only be used to change from one covariant gauge to another at a fixed loop level, but also to predict higher-loop terms from lower-loop ones. However, those predicted terms will all be gauge parameter dependent \cite{LKFT-Loops1,LKFT-Loops2,LKFT-Loops3,LKFT-Loops4}. There have been many efforts to construct the three - point vertex in a way that would ensure the LKFT law for the massless fermion propagator, see, 
for example, \cite{LKFT-Loops1,LKFT-Loops2,LKFT-Loops3,LKFT-Loops4,curtpenn,roberts,adnpenn,pennkiz,enhancedmm,adnanroberts,Aslam-2015}. 

In this contribution, which is based on our recent paper \cite{multiphotonprd}, we consider the less explored case of scalar QED. Applying the worldline formalism, we present an off-shell multiphoton amplitude master formula, as well as a generalization of the LKFT to arbitrary $x$-space amplitudes.
The worldline formalism is based on Feynman's early and well-known representation of the QED S-matrix in terms of first-quantized path integrals \cite{feyn}
whose potential for improving on the efficiency of calculations in QED was recognized only following the work of Strassler in 1992 \cite{strassler-92} (see also \cite{5,18,41}) . 
That work in turn was triggered by an effort by Bern and Kosower \cite{berkos} who found a
novel and efficient way to compute gauge theory amplitudes representing them as the infinite string tension limit of suitable string amplitudes. 
In particular, in this way they obtained a compact generating function for the one-loop (on-shell) N - gluon amplitudes, known as the Bern-Kosower master formula.
The worldline formalism allows one to rederive this master formula in a purely field theoretical approach \cite{strassler-92,18}. 
More recently, two of the present authors have applied the worldline formalism along these lines to recalculate the off-shell three-gluon
vertex~\cite{92}, recuperating the form-factor decomposition originally proposed by Ball and Chiu~\cite{ball-chiu-3gluon} in a way that not only significantly reduces the amount of algebra,  but also allows one to combine the scalar, spinor and gluon loop cases. The superior efficiency of the method becomes even more conspicuous at
the four-gluon level~\cite{98}. 

Concerning the scalar QED case, already in 1996, Daikouji {\em et. al.}~ \cite{dashsu} applied the string-inspired worldline formalism to 
amplitudes in this theory, albeit only in momentum space. Here we follow a similar approach, only that for our present purposes
it will be essential to work in $x$-space as well as in momentum space. 
We will first derive a master formula for nonlinear Compton scattering with an arbitrary number of photons, on-shell and off-shell. 
These amplitudes are becoming relevant these days for laser physics, for a review see \cite{di piazza}. 
Then, taking advantage of the fact that, in the worldline formalism, changes of the gauge parameter can be implemented by total
derivative terms under the path integration, we derive a generalization of the LKFT to arbitrary amplitudes in scalar QED. 
We discuss the perturbative workings of this transformation, 
and finally exemplify both the usefulness of the master formula and of the
generalized LKFT by a recalculation of the one-loop propagator and vertex in an arbitrary covariant gauge.

\section{Master formula for generalized Compton scattering in scalar QED}
\label{masterformula}

In this section we will present a master formula for the scalar propagator to absorb and emit $N$ photons along the way of its propagation from $x'$ to $x$. 
Feynman's path integral representation of the scalar propagator  of mass $m$ in the presence of a background $A(x)$ is
\bear
\Gamma[x,x']=\int_0^\infty dT{\rm e}^{-m^2T}\int_{x(0)=x'}^{x(T)=x}\mathcal{D}x(\tau){\rm e}^{-\int_0^Td\tau[\frac{1}{4}\dot{x}^2+ie\dot{x}\cdot A(x)]+\frac{e^2}{2}\int_0^Td\tau_1\int_0^Td\tau_2\dot{x}_1^\mu\,D_{\mu\nu}(x_1-x_2)\dot{x}_2^\nu}\,.
\label{e1}
\ear
The last term in the exponential gives the virtual photons exchanged along the scalar's trajectory. $D_{\mu\nu} $ is the $x$-space photon propagator, 
which in $D$ dimension and an arbitrary covariant gauge is given by

\bear
D_{\mu\nu}(x) =
\frac{1}{4\pi^{\frac{D}{2}}}
\Big\{\frac{1+\xi}{2}\Gamma\Big(\frac{D}{2}-1\Big)\frac{\delta_{\mu\nu}}{{(x^2)}^{\frac{D}{2}-1}}+(1-\xi)\Gamma\Big(\frac{D}{2}\Big)
\frac{x_{\mu}x_{\nu}}{{(x^2)}^{\frac{D}{2}}}\Big\}\,.
\ear
By choosing $A(x)$ to be a sum of plane waves $A^\mu(x)=\sum_{i,j=1}^N\,\varepsilon_i^\mu{\rm e}^{ik_i\cdot x}$, each external photon 
effectively gets represented by a {\rm vertex operator} 

\bear
V^A_{\rm scal}[k,\varepsilon]\equiv \varepsilon_\mu\int_0^Td\tau\dot{x}^\mu(\tau)\,{\rm e}^{ik\cdot x(\tau)}=\int_0^Td\tau\, {\rm e}^{ik\cdot
x(\tau)+\vaeps\cdot \dot{x}(\tau)}\Big\vert_{{\rm lin}~ \vaeps}\,.
\label{defvertop}
\ear
The path integral is computed by splitting $x^\mu(\tau)$ into a ``background'' part $x^{\mu}_{\rm bg}(\tau)$, which encodes the
boundary conditions, and a fluctuation  part $q^\mu(\tau)$, which has Dirichlet boundary conditions at the endpoints $\tau =0, T$:
 \bear x(\tau)&=&x_{\rm
bg}(\tau)+q(\tau)\non
 x_{\rm
bg}(\tau)&=&x'+\frac{(x-x')\tau}{T}\non
\dot{x}(\tau)&=&\frac{x-x'}{T}+\dot{q}(\tau)\non
 q(0)&=&q(T)=0\,.\non
 \label{split}
 \ear
 The calculation of the path integral then requires only the knowledge of the
free path integral normalization, which is
\bear
\int\mathcal{D}q(\tau)\,{\rm e}^{-\int_0^Td\tau\frac{1}{4}\dot{q}^2} = (4\pi T)^{-\frac{D}{2}}\,,
\ear
and of the two-point correlator, given in \cite{mckeon:ap224,basvan-book} 
\bear
\langle
q^\mu(\tau_1)q^\nu(\tau_2)\rangle=-2\delta^{\mu\nu}\Delta(\tau_1,\tau_2)\,,
\ear
with the worldline Green function
\bear
\Delta(\tau_1,\tau_2)=\frac{\tau_1\tau_2}{T}+\frac{\vert\tau_1-\tau_2\vert}{2}-\frac{\tau_1+\tau_2}{2}\,.
\ear 
The $N$ photon amplitude gets represented as

\bear
 \Gamma[x,x';k_1,\veps1;\cdots;k_N,\veps
N]&=&(-ie)^N\intT\int_{x(0)=x'}^{x(T)=x}\matD x(\tau)\,\eTx
\int_0^T\prod_{i=1}^Nd\t i V^A_{\rm
scal}[k_1,\veps1]\cdots V^A_{\rm scal}[k_N, \veps N]\,.
\nonumber\\
\label{bk-open}
 \ear
 Substituting the vertex operator (\ref{defvertop}) in Eq.~(\ref{bk-open}), and applying the split in Eq.~(\ref{split}), one gets
 
 \bear
\Gamma[x,x';k_1,\veps1;\cdots;k_N,\veps N]&=&(-ie)^N\intT\, {\rm
e}^{-\frac{1}{4T}(x-x')^2}\int_{q(0)=q(T)=0}\matD q(\tau)\,\eTq\non
&\times&\int_0^T\prod_{i=1}^Nd\t i\,{\rm
e}^{\sum_{i=1}^N\big(\veps i\cdot \frac{(x-x')}{T}+\veps
i\cdot\dot{q}(\t i)+ik_i\cdot (x-x')\frac{\t i}{T} +ik_i\cdot
x'+ik_i\cdot q(\t i)\big)}\Big\vert_{{\rm lin}(\veps 1\veps2\cdots
\veps N)}\,. \non
\label{master-open-scalar}
 \ear
After completing the square in the exponential, we obtain the following  tree-level Bern-Kosower-type 
master formula in configuration space,

 \bear \Gamma[x,x';k_1,\veps1;\cdots;k_N,\veps N]
&=&(-ie)^N\intT\, {\rm e}^{-\frac{1}{4T}(x-x')^2}\big(4\pi
T\big)^{-\frac{D}{2}}\non && \hspace{-3.5cm}
\times\int_0^T\prod_{i=1}^Nd\t i\,{\rm e}^{\sum_{i=1}^N\big(\veps
i\cdot \frac{(x-x')}{T}+ik_i\cdot (x-x')\frac{\t i}{T}+ik_i\cdot
x'\big)}\, {\rm e}^{\sum_{i,j=1}^N\big[\Delta_{ij}\kk
ij-2i\ddel_{ij}\veps i\cdot k_j-\ddeld_{ij}\epseps
ij\big]}\Big\vert_{{\rm lin}(\veps 1\veps2\cdots \veps N)}\,.
\label{bk-like-x}
 \ear
 After Fourier transforming to momentum space one gets the somewhat more compact form 
 
\bear
&&\Gamma[p;p';k_1,\veps1;\cdots;k_N,\veps
N]=(-ie)^N(2\pi)^D\delta^D\Big(p+ p' +\sum_{i=1}^N
k_i\Big)\int_0^\infty dT\, {\rm e}^{-T(m^2+p^2)}\non
&\times&\int_0^T\prod_{i=1}^{N} d\t i\, {\rm
e}^{\sum_{i=1}^N(-2k_i\cdot p\t i+2i\veps i\cdot
p)+\sum_{i,j=1}^N\big[(\frac{\vert \t i-\t j\vert}{2}-\frac{\t i+\t j}{2})\kk
ij-i({\rm sign}(\t i-\t j)-1)\epsk ij+\delta(\t i-\t
j)\epseps ij\big]}\Big\vert_{{\rm lin}(\veps 1\veps2\cdots \veps
N)}\,.\non
  \label{master-bk-open}
\ear
This is our final representation of the multiphoton amplitude in momentum space, see \cite{multiphotonprd} for more details. It is important to mention that it gives the untruncated propagator, including the final scalar propagators on both ends. On-shell it corresponds to multi-photon Compton scattering, while off-shell it can be used for constructing higher-loop amplitudes by sewing. 

\section{Generalization of the LKF transformation}

Coming to the issue of gauge-parameter dependence, let us first consider external photons. 
A gauge transformation of the $i$-th external photon 
\bear
\varepsilon_i\rightarrow \varepsilon_i+\xi k_i
\ear
 changes its vertex operator by boundary terms:

 \bear
 V_{\rm scal}[\veps i,k_i]=\int_0^T d\tau_i \vaeps_{i\mu}
\dot{x}_i^\mu\, {\rm e}^{ik_i\cdot x(\tau_i)}\rightarrow V_{\rm
scal}[\veps i,k_i]-i\xi\int_0^T
\frac{\partial}{\partial\tau_i}\,{\rm e}^{ik_i\cdot
x(\tau_i)}=V_{\rm scal}[\veps i,k_i]-i\xi\Big({\rm e}^{ik_i\cdot
x}-{\rm e}^{ik_i\cdot x'}\Big)\,.
 \ear
This is just the QED Ward identity, which we need not discuss further. 
More interesting is the gauge transformation of internal photons, given by the last term in the exponential of Eq. (\ref{e1}).
Its integrand can be written as 
  
\bear
 \frac{1}{4\pi^{\frac{D}{2}}}
 \biggl\lbrack
 \Gamma\Big(\frac{D}{2}-1\Big)\frac{\dot{x}_1\cdot\dot{x}_2}{[(x_1-x_2)^2]^{\frac{D}{2}-1}}-\frac{1-\xi}{4}
 \Gamma\Big(\frac{D}{2}-2\Big)\frac{\partial}{\partial\tau_1}\frac{\partial}{\partial\tau_2}[(x_1-x_2)^2]^{2-\frac{D}{2}}
 \biggr\rbrack
 \,.
 \label{gauge-D}
 \ear
 This shows that a change of the gauge parameter $\xi$ by $\Delta\xi$ changes the integrand only by a (double) total derivative:
 
 \bear
 \Delta\xi  \frac{e^2}{32\pi^{\frac{D}{2}}}
\Gamma\Big(\frac{D}{2}-2\Big)
\int_0^Td\tau_1\int_0^Td\tau_2
\frac{\partial}{\partial\tau_1}\frac{\partial}{\partial\tau_2}[(x_1-x_2)^2]^{2-\frac{D}{2}}\,.
\label{deltaSi}
 \ear
Thus if one or both ends of the photon sit on a closed loop this term vanishes. 
Therefore the gauge transformation properties of any amplitude in scalar QED are determined by the photons exchanged between two scalar lines, or along one scalar line. 
Thus we can discard not only all external photons but also all closed scalar loops, so that it suffices to study the quenched $2n$ scalar amplitude.
This amplitude can be written as

\bear A^{\rm qu}(x_1,\ldots,x_n;x_1',\ldots,x_n'\vert\xi)=
\sum_{\pi\in S_n} A^{\rm
qu}_{\pi}(x_1,\ldots,x_n;x'_{\pi(1)},\ldots,x'_{\pi(n)}\vert\xi)\,,
 \ear
 where in the partial amplitude $A^{\rm qu}_{\pi}(x_1,\ldots,x_n;x'_{\pi(1)},\ldots,x'_{\pi(n)}\vert\xi)$
it is understood that the line ending at $x_i$ starts at
$x'_{\pi(i)}$. The worldline representation of this amplitude at the quenched level is \cite{feyn}: 

\bear
A^{\rm qu}_{\pi}(x_1,\ldots,x_n;x'_{\pi(1)},\ldots,x'_{\pi(n)}\vert\xi) =
\prod_{l=1}^n
\int_0^\infty dT_l\, {\rm e}^{-m^2T_l}\,
  \int_{x_l(0)=x'_{\pi(l)}}^{x_l(T_l)=x_l}\matD x_l(\tau_l)
 \, {\rm e}^{-\sum_{l=1}^n S_0^{(l)} -\sum_{k,l=1}^n S_{i\pi}^{(k,l)}}\,.
\label{Aquenched}
 \ear
 Here, 
 \bear
  S_ 0^{(l)} =  \int_0^{T_l} d\tau_l \frac{1}{4}\dot{x_l}^2\,,
  \ear
 is the free worldline Lagrangian for the path integral representing line $l$,  and 
\bear
S_{i\pi}^{(k,l)}= \frac{e^2}{2}
\int_0^{T_k} d\tau_{k}\int_0^{T_l} d\tau_{l}\, \dot{x}^{\mu}_k D_{\mu\nu}(x_k-x_l)\dot{x}^{\nu}_l\,,
\ear
generates all the photons connecting lines $k$ and $l$. 
Thus, after a gauge change,

\bear
A^{\rm qu}_{\pi}(x_1,\ldots,x_n;x'_{\pi(1)},\ldots,x'_{\pi(n)}\vert\xi + \Delta\xi) =
\prod_{l=1}^n
\int_0^\infty dT_l\, {\rm e}^{-m^2T_l}\,
  \int_{x_l(0)=x'_{\pi(l)}}^{x_l(T_l)=x_l}\matD x_l(\tau_l)
 \, {\rm e}^{-\sum_{l=1}^n S_0^{(l)} -\sum_{k,l=1}^n \bigl(S_{i\pi}^{(k,l)}+ \Delta_{\xi}S_{i\pi}^{(k,l)}\bigr)}\,,\non
\label{Aquenchedtransformed}
 \ear where, from Eq.~(\ref{deltaSi}), we have
 \bear \Delta_\xi  S_{i\pi}^{(k,l)} &=& \Delta\xi  \frac{e^2\Gamma\big(\frac{D}{2}-2\big)}{32\pi^{\frac{D}{2}}}
 \biggl\lbrace
\bigl[(x_k-x_l)^2\bigr]^{2-D/2}-\bigl[(x_k-x'_{\pi(l)})^2\bigr]^{2-D/2}
-\bigl[(x'_{\pi(k)}-x_l)^2\bigr]^{2-D/2}+\bigl[(x'_{\pi(k)}-x'_{\pi(l)})^2\bigr]^{2-D/2}
\biggr\rbrace\,. \non
 \ear 
Since $\Delta\xi$ depends only on the
endpoints of the scalar trajectories we can rewrite (\ref{Aquenchedtransformed}) as 
\bear
A^{\rm qu}_{\pi}(x_1,\ldots,x_n;x'_{\pi(1)},\ldots,x'_{\pi(n)}\vert\xi + \Delta\xi) =
T_{\pi}
A^{\rm qu}_{\pi}(x_1,\ldots,x_n;x'_{\pi(1)},\ldots,x'_{\pi(n)}\vert\xi)\,,
 \label{central}
\ear
with 
\bear
T_{\pi} \equiv \prod_{k,l=1}^N  \, {\rm e}^{- \Delta_{\xi}S_{i\pi}^{(k,l)}}\,.
\ear
 
This is an exact $D$-dimensional result. When
using dimensional regularization around  $D=4$, one has to
take into account that the full non-perturbative $A^{\rm qu}$ in
scalar QED has poles in $\epsilon$ to arbitrary order, so that
also the prefactor $T_{\pi}$, although regular, needs to be kept
to all orders. Here we will consider only the leading constant
term of this prefactor. Thus, we compute
\bear {\rm lim}_{D\to 4} \, {\rm e}^{-
\Delta_{\xi}S_{i\pi}^{(k,l)}} =  \Bigl(r^{(k,l)}_{\pi}\Bigr)^c\,,
\label{lim}
 \ear
 where we have introduced the constant $c \equiv \Delta\xi  \frac{e^2}{32\pi^2}$
and the conformal cross ratio $ r^{(k,l)}_{\pi}$ associated to the four endpoints of the lines $k$ and $l$,
\bear
 r^{(k,l)}_{\pi}
\equiv \frac{ (x_k-x_l)^2 (x'_{\pi(k)}-x'_{\pi(l)})^2 } {
(x'_{\pi(k)}-x_l)^2 (x_k-x'_{\pi(l)})^2 }\,.
 \ear
 Thus, at the leading order, the prefactor turns into
\bear
T_{\pi} = \biggl(\prod_{k,l=1}^N r^{(k,l)}_{\pi}\biggr)^c +
O(\epsilon)\,.
\ear
 We note that for the case of a
single propagator, $s=k=l=1$, $T_{\pi}$ degenerates into
\bear T =
\biggl\lbrack\frac{(x-x)^2(x'-x')^2}{((x-x')^2)^2}\biggr\rbrack^c\,.
\ear 
Replacing the vanishing numerator $(x-x)^2(x'-x')^2$ by the cutoff  $(x_{\rm min}^2)^2$, and
$\Delta\xi$ by $\xi$, we recuperate the original LKFT

\bear
S_F(x;\xi)=S_F(x;0) \Big(\frac{x^2}{x_{\rm min}^2}\Big)^{-e^2\xi/(4\pi)^2}\,.
\ear

\section{The generalized LKFT in perturbation theory}
\label{lkftp}
\begin{figure}[h]
  \centering
    \includegraphics[width=0.3\textwidth]{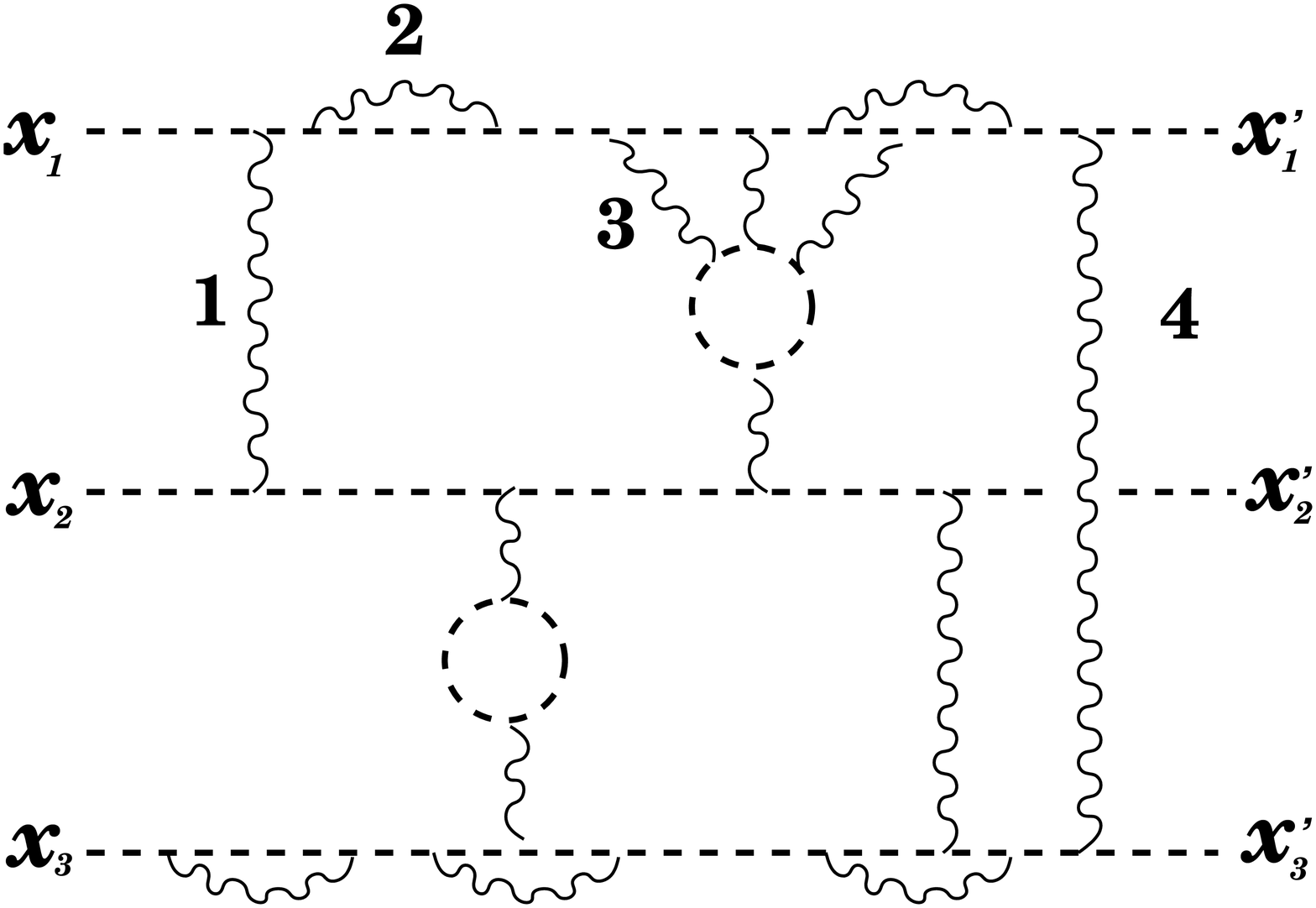}
      \caption{Feynman diagram representing six - scalar amplitude at twelve loops.}
      \label{fig-LKF0}
\end{figure}

In this section we will show by an example how the non-perturbative gauge transformation presented in previous section and in Eq.~(\ref{central}) works in perturbation theory.
Consider the twelve-loop contribution to the scalar six-point function shown in FIG. \ref{fig-LKF0}

\begin{figure}[H]
\centering
\subfigure[Gauge transformation of photon $1$.]{%
\includegraphics[width=0.25\textwidth]{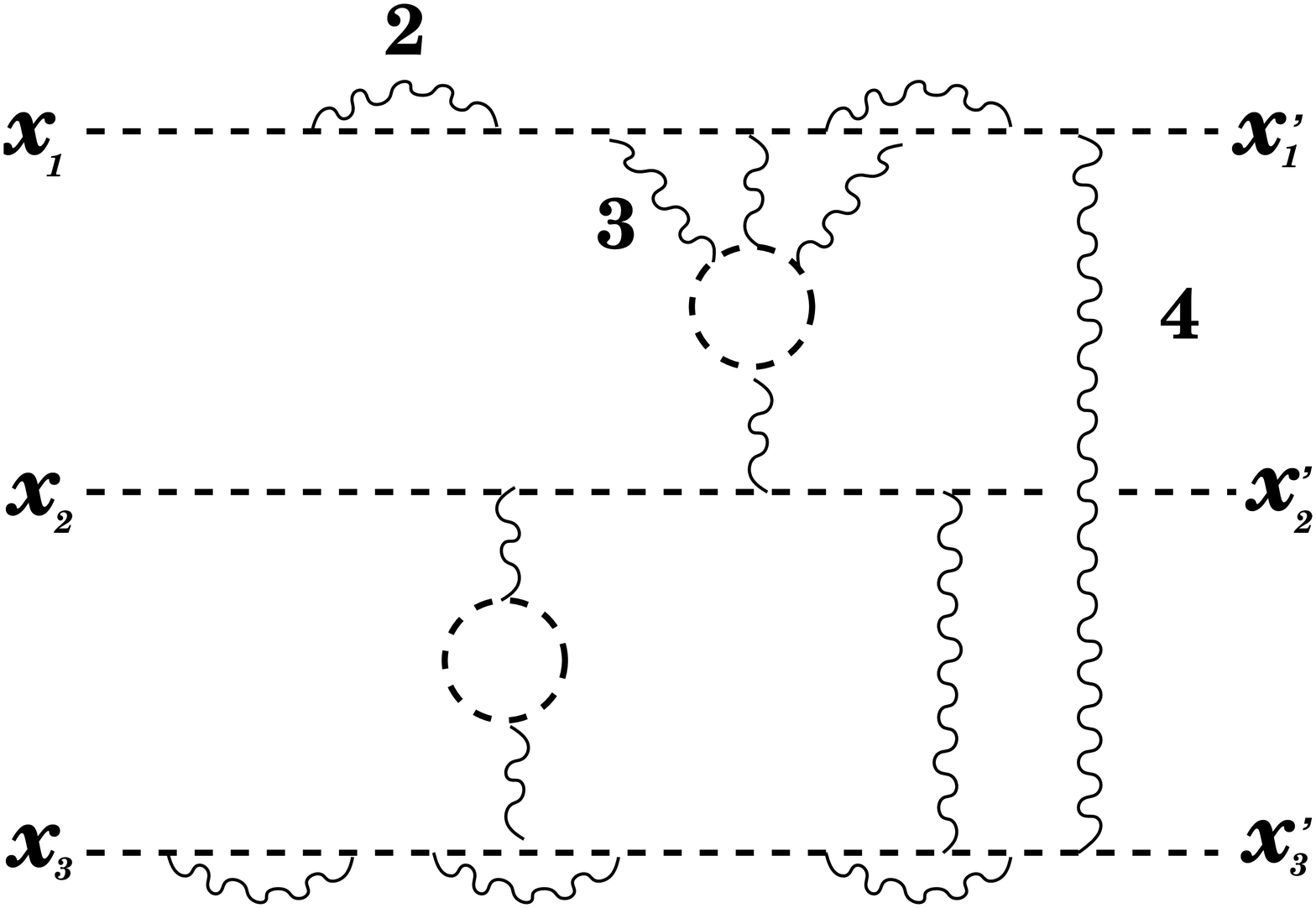}
\label{LKF1}}
\quad
\subfigure[Gauge transformation of photon $2$.]{%
\includegraphics[width=0.25\textwidth]{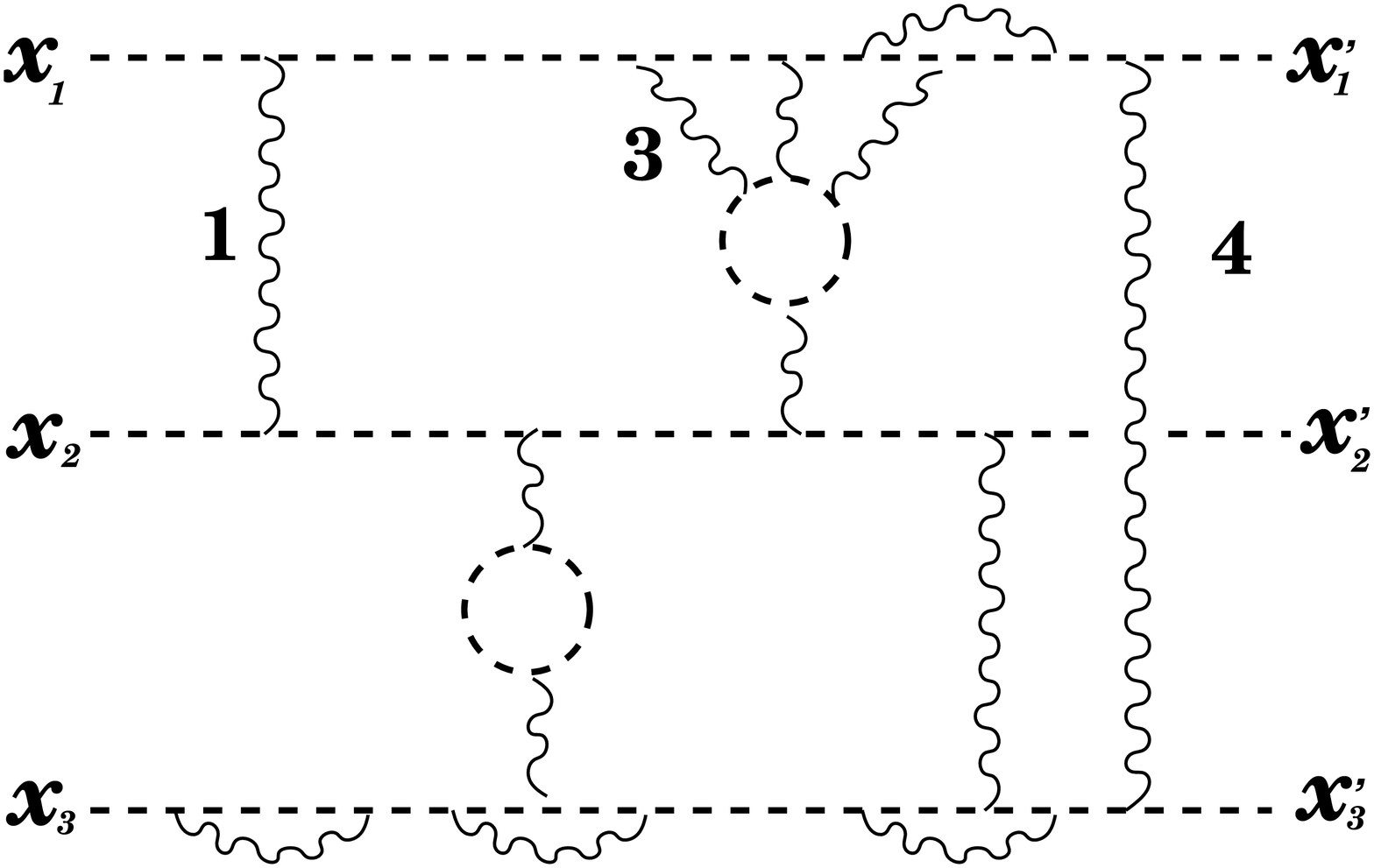}
\label{LKF2}}
\subfigure[Gauge transformation of photon $4$.]{%
\includegraphics[width=0.25\textwidth]{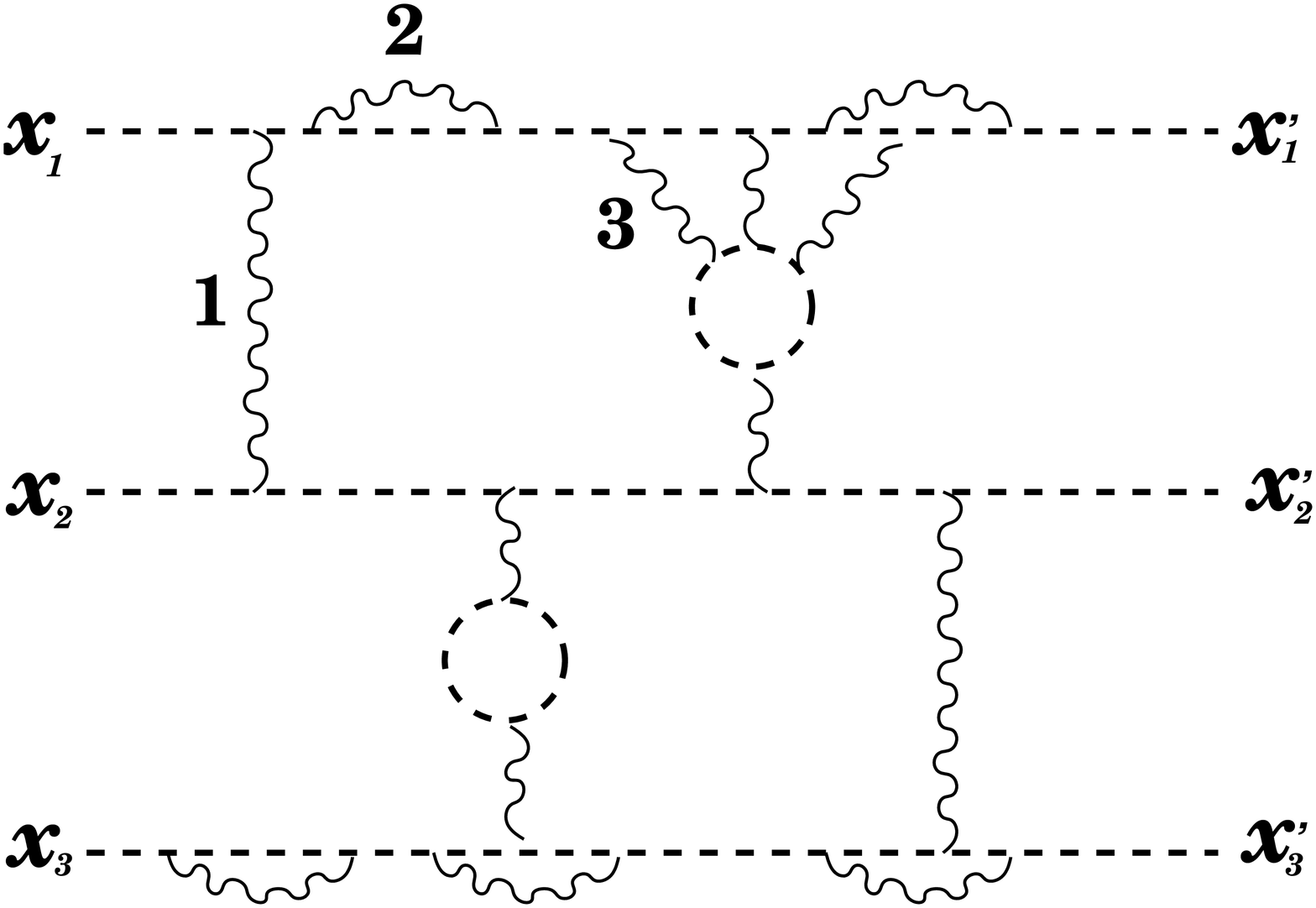}
\label{LKF4}}
\quad
\subfigure[Simultaneous gauge transformation of photons $1$ and $2$.]{%
\includegraphics[width=0.25\textwidth]{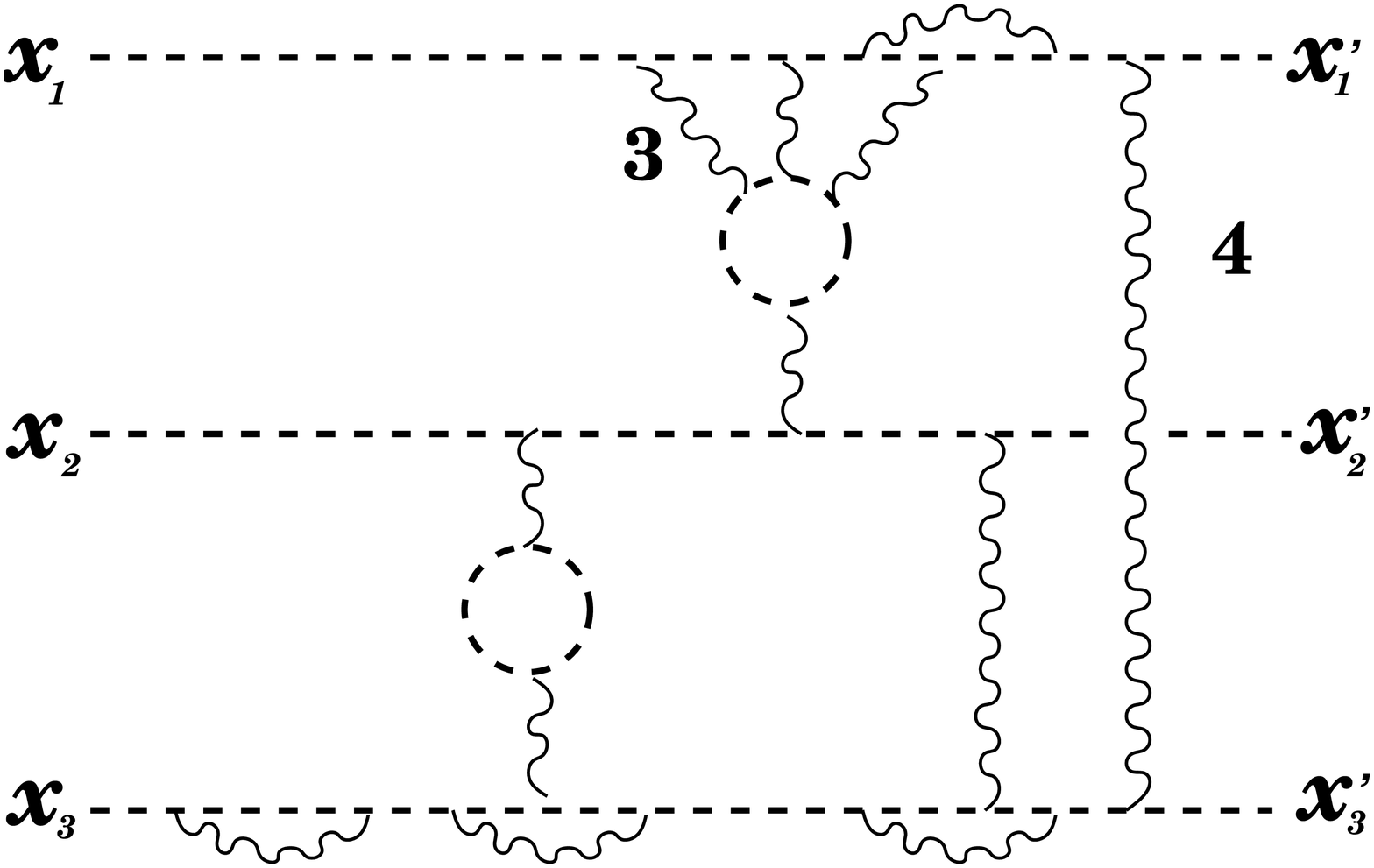}
\label{LKF12}}
\quad
\subfigure[Simultaneous gauge transformation of photons $1$, $2$ and $4$.]{%
\includegraphics[width=0.25\textwidth]{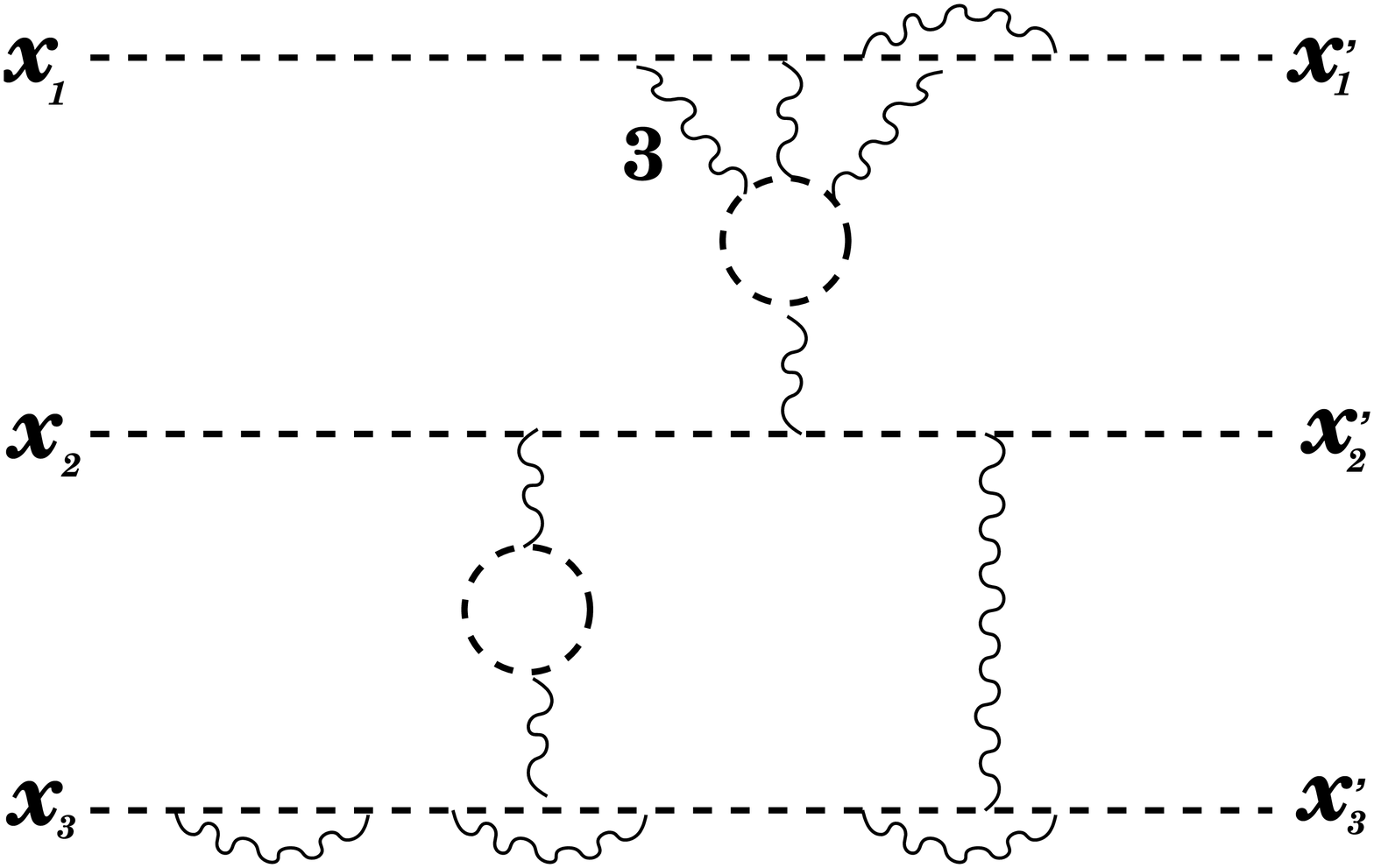}
\label{LKF124}}
\caption{Gauge transformation of internal photons.}
\label{generalLKFT-x}
\end{figure}

The change of the gauge should be considered for all photons (except the ones ending on a loop). For each photon its gauge transformation leaves a diagram without the transformed photon. The gauge transformation of the whole set diagrams is called $\Delta_\xi$ Fig. \ref{fig-LKF0} which is written as   
\bear
 \Delta_{\xi}\,{\rm Fig}\,\ref{fig-LKF0} &=& \big(- 2\Delta_{\xi}S_{i\pi}^{(1,2)})\,{\rm Fig}\,\ref{LKF1}
+\big(- \Delta_{\xi}S_{i\pi}^{(1,1)})\,{\rm Fig}\,\ref{LKF2}
+\big(-2 \Delta_{\xi}S_{i\pi}^{(1,3)})\,{\rm Fig}\,\ref{LKF4}+\cdots \non
&&+\big(-2 \Delta_{\xi}S_{i\pi}^{(1,2)}\big)\big(- \Delta_{\xi}S_{i\pi}^{(1,1)}\big)\,{\rm Fig}\,\ref{LKF12}+\cdots\non
&&+\big(-2 \Delta_{\xi}S_{i\pi}^{(1,2)}\big)\big(- \Delta_{\xi}S_{i\pi}^{(1,1)}\big)\big(-2 \Delta_{\xi}S_{i\pi}^{(1,3)}\big){\rm Fig}\,\ref{LKF124}+\cdots\non &&+\cdots \,.\nonumber
 \label{KLFT-complicated}
 \ear
 
\section{THE ONE-LOOP PROPAGATOR AND VERTEX IN AN ARBITRARY COVARIANT GAUGE}

Mathematically the LKFT is more difficult to implement in momentum than in configuration space, such that there is little hope
for a closed-form non-perturbative formula such as (\ref{central}) in momentum space. However, it remains true that gauge paramter changes
are implemented by total derivative terms, and we will now show for the example of the one-loop propagator how
this can be used to easily obtain an amplitude in a general covariant gauge, once it is known in some particular gauge.
For this purpose, if we take $N=2$ from our master formula  and sew the two photons in arbitrary covariant gauge using

\bear
\varepsilon_1^\mu\varepsilon_2^\nu\rightarrow
\frac{\delta^{\mu\nu}q^2-(1-\xi)q^\mu q^\nu}{q^4}\,,
\label{covariant gauge}
 \ear
 we get
  \bear
 \Gamma_{\rm propagator}(p)&=&-e^2(m^2+p^2)^2\int_0^\infty dT T^2\, {\rm
e}^{-T(m^2+p^2)}\int_0^1du_1\int_0^{u_1}du_2\int
\frac{d^Dq}{(2\pi)^D}\non &\times& 
(2p_{\mu}+q_{\mu})(2p_{\nu}+q_{\nu})
\Big[\frac{\delta^{\mu\nu}}{q^2}+(\xi-1)\frac{q^\mu
q^\nu}{q^4}\Big]\, {\rm e}^{-T(u_1-u_2)(q^2+2p\cdot q)}
\,.\non
\ear
By using the fact that 

\bear
 &&
 (2p_{\mu}+q_{\mu})(2p_{\nu}+q_{\nu})
 (\xi-1)\frac{q^\mu q^\nu}{q^4}\,{\rm e}^{-T(u_1-u_2)(q^2+2p\cdot q)}=-\frac{(\xi-1)}{T^2q^4}\frac{\partial^2}{\partial u_1\partial u_2}\,{\rm e}^{-T(u_1-u_2)(q^2+2p\cdot q)}
 \ear
 one finally gets the following result for one-loop correction to the propagator in any covariant gauge 
 \bear
 \Gamma_{\rm propagator}(p)
&=&\frac{e^2}{m^2}\Big(\frac{m^2}{4\pi}\Big)^{\frac{D}{2}}
 \Gamma\Big(1-\frac{D}{2}\Big)\bigg\{1-2\frac{(m^2-p^2)}{m^2}\,_2F_1\Big(2-\frac{D}{2},1;\frac{D}{2};-\frac{p^2}{m^2}\Big)\non
&&\hspace{3cm}+(1-\xi)\,
\frac{(m^2+p^2)^2}{m^4}\,_2F_1\Big(3-\frac{D}{2},2;\frac{D}{2};-\frac{p^2}{m^2}\Big)\bigg\}\,.\non
\label{p-s-propagator}
 \ear
Similarly, and much more non-trivially, one can use the same strategy to obtain the one-loop correction to the vertex in an arbitrary covariant gauge
from the one in, say, Feynman gauge. Here, however, for lack of space we must refer the reader to \cite{multiphotonprd}.

\section{Conclusion}
In this contribution we have applied the worldline formalism to obtain a Bern-Kosower type master formula for the scalar propagator dressed with any number of photons,
on- and off-shell. We have used this formula to find the one-loop scalar propagator in any covariant gauge. 
In $x$-space, the implementation of changes of the gauge parameter through total derivatives has allowed us to obtain, in a very simple way, an explicit non-perturbative formula for the effect of such a gauge parameter change on an arbitrary amplitude to all loop orders. This formula generalizes the LKFT and contains it as a special case. At leading order in the $\epsilon$~-~expansion it can be written in terms of conformal cross ratios. We also have illustrated with an example how this non-perturbative transformation works diagrammatically in perturbation theory. One possible extension of our results would be to spinor case which is under study. A non-abelian version of the master formula 
has been obtained recently in \cite{ahbaco-colorful}.

\end{document}